\documentclass[12pt]{iopart}

\usepackage{graphicx}

\begin{document}

\title{Quantum interference effects on the noise power in the CNT/aGNR/CNT junction}

\author{A Ahmadi Fouladi}
\address{Department of Physics, Sari Branch, Islamic Azad University, Sari, Iran}
\ead{a.ahmadifouladi@iausari.ac.ir}

\author{Javad Vahedi}
\address{Department of Physics, Sari Branch, Islamic Azad University, Sari, Iran}
\ead{javahedi@iausari.ac.ir}

\begin{abstract}
Based on tight-binding model and a generalized Green's function method in Landauer-B\"uttiker formalism, the effects of quantum interference (QI) on the noise power and Fano factor of an armchair graphene nanoribbon (aGNR) sandwiched between infinite single wall carbon nanotube (SWCNT) as a CNT/aGNR/CNT system are numerically investigated. In this work, changing the aGNR to CNT electrodes contact positions and applying the magnetic field as two sources of QI are considered. We have found different Fano-resonance and anti-resonance peaks on the transmission probability in the presence of QI sources that show profound effects on the current-voltage characteristics and noise power. Our results also show that the shot noise characteristic, either in the Poisson limit ($F=1$) or sub-Poisson limit ($F<1$), and also maximum value of the Fano factor strongly depend on the aGNR to CNT electrodes contact positions and the magnetic field strength.
These results can be useful for designing the future nano-electronic devices.
\end{abstract}

\maketitle

\section{Introduction}
\par
Recently, the electronic transport properties through the graphene nanoribbons (GNRs) have attracted a great deal of interest due to the potential
applications in nano electronic devices \cite{1,2,3,4}.
So devices have designed in such a way that a GNR is sandwiched between two metalic or organic electrodes.
The tiny dimension, strike and profound physical properties of these structures make them prospective candidates for future generation
of nano-devices in electronic engineering.
 \par
One of the prominent factors in the transport properties through the nano scale structures is the quantum interference (QI) effects
which attracted much more attention in the past few years \cite{5,6}.
Such effects occur when electron waves pass through nano junctions in a phase-coherent fashion and depend crucially on symmetry.
The relevance of QI effects has been first observed in the context of waveguides for semiconductor
nanostructures \cite{7,8,9}, but it has also been recognized early on theoretically \cite{10} and experimentally\cite{11,12}.
QI effects in graphene subjected to a perpendicular magnetic field, such as current revivals\cite{13}
or Aharonov–Bohm conductance oscillations in ring-shaped devices have also been investigated \cite{14,15,16,17,18,19}.
Munarriz et al\cite{19} investigated QI effects in a type of QI device based on a graphene nanoring in which all edges
are of the same type and they have shown that the superposition of the electron wavefunction propagating from the source to
 the drain along the two arms of the nanoring gives rise to
interesting interference effects.
 \par
In the other hand, to obtain  additional information about charge carriers in transport phenomena
 one can study noise power of the current fluctuations (of thermal or quantum origin) which give several key ideas for fabrication of the
 efficient nano-electronic devices. In order to design nanometer junctions with the lowest
possible noise power, it is necessary to have an efficient algorithm to calculate the noise power current fluctuations. In a review work, Blanter and B\"uttiker have shown clearly and elaborately how the lowest
possible noise power of the current fluctuations can be determined in a two-terminal conductor \cite{20}.
 Shot noise (steady-state current fluctuation) which is a consequence of  the granularity of charge  has been the subject
 of intensive theoretical and experimental research \cite{20,21,22,23,24,25,26}. In addition, relying on  the theory of
 shot noise \cite{20} one may define a Fano factor (F) which presents information about  electron correlations of the system.
 \par
In the present paper, motivated by the significantly improved switching characteristics of the short
organic field effect transistors (OFET) with carbon nanotube (CNT) electrodes compared to the metallic
ones \cite{27,28,29,30,31,32,33},
 we have numerically investigated the effects of QI on the noise power and Fano factor of the armchair graphene nanoribbon (aGNR) sandwiched between an armchair $(l,l)$ single-walled carbon nanotube electrodes (CNT/aGNR/CNT system) as an all carbon base nano-junction.
Our calculations based on tight-binding model and a generalized Green's function method in Landauer-B\"uttiker formalism.
In this work, changing the aGNR to CNT electrodes contact positions and applying the magnetic field as two sources of QI are considered.
\par
The paper is organized as follows. The model, Hamiltonian and formalism are given in section II.
In section III, we present the results of the numerical calculations. Finally, conclusion is given in section IV.

\begin{figure}
\centerline{\includegraphics[width=0.7\columnwidth]{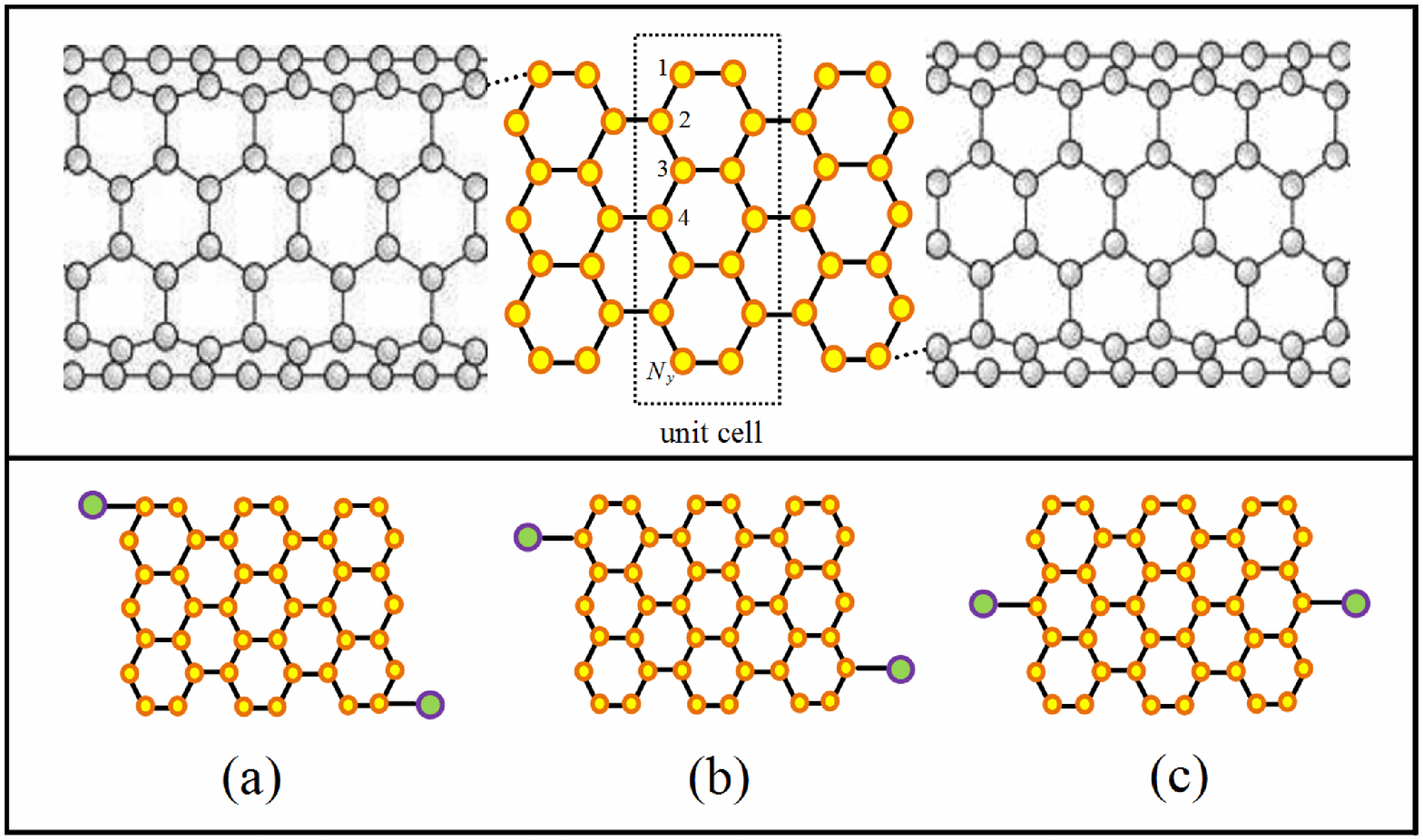}}
\caption{(color online.) The upper panel shows CNT/aGNR/CNT system. The bottom panel represents the different location of coupling
 between aGNR and two left and right CNT electrodes which are labeled as $a$, $b$ and $c$.}
\label{p1}
\end{figure}
\section{Computational scheme}

We start with Fig.(\ref{p1}), \textbf{\emph{where an aGNR lattice with $N_{y}=7$ and three unit cells}}, subject to a perpendicular magnetic
field is attached to two semi-infinite armchair (5,5) single-walled CNT electrodes. These electrodes are characterized by the
electrochemical potentials $\mu_{L}$ and $\mu_{R}$, respectively, under the non-equilibrium condition when an external bias
voltage is applied. The Hamiltonian representing the entire system can be written as a sum of four terms,
\begin{equation}
\emph{H}=H_{L}+H_{GNR}+H_{R}+H_{C},
\label{e1}
\end{equation}
where $H_{L(R)}$ represents the Hamiltonian of the left (right) CNT electrode which is described within
the tight binding approximation and is written as
\begin{eqnarray}
\emph{H}_{\gamma}=\sum_{i_{\gamma}}\{\varepsilon_{0} c_{i_{\gamma}}^{\dag}c_{i_{\gamma}}\
     -t_{0}(c_{i_{\gamma}}^{\dag}c_{i_{\gamma}+1}+c_{i_{\gamma}+1}^{\dag}c_{i_{\gamma}})\}.
\label{e2}
\end{eqnarray}
Here $c_{i_{\gamma}}^{\dag}(c_{i_{\gamma}})$ denotes the creation (annihilation) operator of an electron at site
$i$ in the electrode $\gamma(=L or R)$. $\varepsilon_{0}$ and $t_{0}$ are the on-site energy and the nearest-neighbor hopping
integral, respectively. In the absence of CNT electrodes, the Hamiltonian of aGNR may be described by tight-binding model within
the non-interacting picture like,
\begin{eqnarray}
\emph{H}_{GNR}=\sum_{{n}}\varepsilon_{n}d_{{n}}^{\dag}d_{{n}}-\sum_{{<n,m>}}\tilde{t}_{n,m}(d_{{n}}^{\dag}d_{{m}}+H.c.).\
\end{eqnarray}
The index $n$ runs over the $\pi$-orbitals of the carbon atom on the GNR. The operator $d_{{n}}^{\dag}(d_{{n}})$ creates
(annihilates) an electron at site $n$ on the aGNR. $\varepsilon_{n}$ is the on-site energy of a carbon atom. $\tilde{t}_{n,m}$ corresponds to the nearest-neighbor hopping integral between the carbon atoms in the presence of a perpendicular magnetic field. The effect of magnetic
field $\vec{B}(=\vec{\nabla}\times \vec{A})$ is incorporated in the hopping term $\tilde{t}_{n,m}$ through the Peierl's
phase factor and for a chosen gauge field it becomes,
\begin{equation}
\tilde{t}_{n,m}=t e^{-\frac{i2\pi}{\phi_0}\int_{r_n}^{r_m}\vec{A}\cdot \vec{dl}}
\label{e4}
\end{equation}
where, t gives the nearest-neighbor hopping integral in the absence of magnetic field and $\phi_0 (= 2h/e)$ is the elementary quantum flux.
Finally in Eq.(\ref{e1}), $H_C$ denotes the coupling between the aGNR and CNT electrodes and takes the form;
\begin{equation}
H_C=\sum_{n,i} t_{c(n,i)} (c_{i}^{\dag}d_{i}+H.c.),
\label{e5}
\end{equation}
where the matrix elements $t_{c(n,i)}$ represent the coupling strength between the GNR, and CNT electrodes are taken to be $t_{c}$.
The Green's function of the CNT/aGNR/CNT junction can be written as
\begin{equation}
G(E)=\big[E1-H_{GNR}-\Sigma_{L}(E)-\Sigma_{R}(E)\big]^{-1},
\label{e6}
\end{equation}
where 1 stands for the identity matrix and $\Sigma_{L}(\Sigma_{R})$ is the self-energy matrix resulting from the coupling of
aGNR to the left (right)CNT electrode and given by
\begin{equation}
\Sigma_{\gamma}(E)=\tau_{m,\gamma}\, g\,\tau_{\gamma,m},
\label{e7}
\end{equation}
where $\tau_{m,\gamma}$ is the coupling matrix and it will be non-zero
only for the adjacent points in the aGNR and the CNT electrode. $g$ is the Green's function of the semi-infinite isolated electrodes and given by \cite{27}
\begin{equation}
g_{n_y,n'_y}(E)=\frac {1}{2l}\sum^{2l}_{j=1} \varphi_j(n_y) \tilde G^{j}(E) \varphi^*_j(n'_y),
\label{e8}
\end{equation}
where $\varphi_j(n_y)=\exp (ik^j_y n_y a)$ , with $k^j_ya=\pi jl$ and $1\leq j\leq 2l$ . We assume the $z$ direction to be parallel to the tube
and $y$ to be the finite transverse coordinate. The surface Green's function of CNT electrodes are given by \cite{27}
\begin{equation}
\tilde G^{j}(E)=\frac {E-\varepsilon_{0}}{4{t_0}^2} \Bigg (1+i \frac{\sin(q^{j}_{\beta_ \ast} a)}{\sqrt{(\frac {E-\varepsilon_{0}}{2{t_0}})^2-\sin^2({\frac {\pi j}{l}})}}\Bigg)\,\, ,
\label{e9}
\end{equation}
where ${\beta_ \ast}=sign(E-\varepsilon_{0})$ and
\begin{equation}
\cos(q^{j}_{\beta_ \ast} a))=- \frac {1}{2} \cos({\frac {\pi j}{l}})-\frac {\beta}{2} \sqrt{(\frac {E-\varepsilon_{0}}{2{t_0}})^2-\sin^2({\frac {\pi j}{l}})}.
\label{e10}
\end{equation}
In general, there are $N=2l$ atomic positions over the interfacial end-atoms of the nanotube. For an armchair $(l,l)$ single-walled carbon nanotube (SWNT)
topology imposed, the number of carbon sites at the interface are $2l$.
The transmission probability $T(E)$ can be expressed in terms of the Green's function of the aGNR and the coupling of
the aGNR with two CNT electrodes by expression
\begin{equation}
T(E)=Tr(\Gamma_L\, G^r \Gamma_R G^a).
\label{e11}
\end{equation}

Where $G^r$ and $G^a$ are respectively the retarded and advanced Green's function.
The broadening matrix $\Gamma_{L,R}$ is defined as the imaginary part of the self-energy;
\begin{equation}
\Gamma_{L,R}=-2 Im(\Sigma_{L,R}).
\label{e12}
\end{equation}
Transmission function tells us the rate at which electrons transmit from the left to the right electrode by propagating through
the aGNR. In the low bias limit, the current as a function of the applied voltage $V$ can be calculated in the Landauer-B\"uttiker formula
based on the non-equilibrium Green'd function method \cite{34};
\begin{equation}
I{(V)}=\frac{2e}{h}\int_{-\infty}^{+\infty}T(E)\big[f_{L}-f_{R}\big] dE.
\label{e13}
\end{equation}
where $f_{L(R)}=f(E-\mu_{L(R)})$ is the Fermi distribution function at the left (right)
electrode with chemical potential $\mu_{L(R)}=E_{F}\pm\frac{eV}{2}$ and Fermi energy $E_{F}$.
For the sake of simplicity, here we assume that the total voltage is dropped across aGNR interface
and this assumption does not eminently affect the qualitative aspects of current-voltage ($I-V$) characteristics. In fact
the electric field inside the aGNR, particularly for narrow aGNR, seems to have a negligible effect on
the $I-V$ characteristics. On the contrary, for quite larger aGNR and higher bias voltages, the electric field inside
the aGNR may play a more remarkable role depending on the structure and size of the aGNR \cite{35}, but yet the effect is
very small. The noise power of the current fluctuations provides additional insight into the quantum transport
properties and statistics of the electron transfer. The zero-frequency noise power of the current fluctuations is
given through the following relation \cite{20};
\begin{eqnarray}
S(E)&=&\frac{4e^2}{h}\int_{-\infty}^{+\infty}\big[\,T(E)\{f_{L}(1-f_{L})+f_{R}(1-f_{R})\}\nonumber \\
    &+& T(E)\{1-T(E)\}(f_{L}-f_{R})^2\,\big]\, dE.
\label{e14}
\end{eqnarray}
The first two terms describe equilibrium noise originated from thermal fluctuation in the left and right electrodes, respectively.
These contributions vanish at zero temperature. The last term describe shot noise term due to the discrete nature of the electron
transport. A suitable measure of the magnitude of the shot noise is Fano factor contains additional considerable information about
electron correlation in the system \cite{20} and is given as $F=\frac{S}{2eI}$ with $e$ being the elementary quantum of charge.
The full shot noise is characterized by $F=1$. In this case the shot noise achieves the Poisson limit where there is no correlation
between the charge carriers. For $F<1$ the sub-Poisson value of shot noise is achieved where electron correlations reduce the level of
current fluctuations below the Poisson limit.

\section{Result and discussion}
Here, we represent the results of the numerical calculations based on the formalism described in section 2.
We set the carbon on-site energy and hopping energies of aGNR as $\varepsilon_n=0$ and $t=2.71 eV$, respectively.
The tight-binding parameters for CNT electrodes are chosen to be $\varepsilon_{0}=0\,$, $t_{0}=2.71\,eV$.
As a reference energy, the Fermi energy of CNT electrodes is set $E_F=0$. For the sake of simplicity we consider
elementary quantum flux and lattice constant in aGNR as $\phi_0=1$  and $a=1$, respectively.
In addition, we set $t_c=0.5\,eV$ and $T=4\,K$.
\begin{figure}
\centerline{\includegraphics[width=0.6\columnwidth]{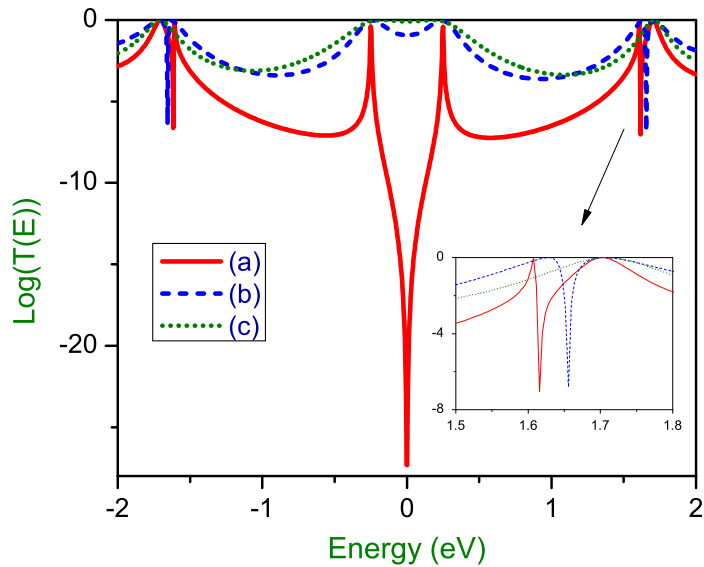}}
\caption{(color online.) The logarithmic scale of transmission probability as a function of the injecting electron
energy $E$ of the CNT/aGNR/CNT junction in the absence of magnetic field for different $a, b$ and $c$ cases.}
\label{p2}
\end{figure}

\begin{figure}
\centerline{\includegraphics[width=0.55\columnwidth]{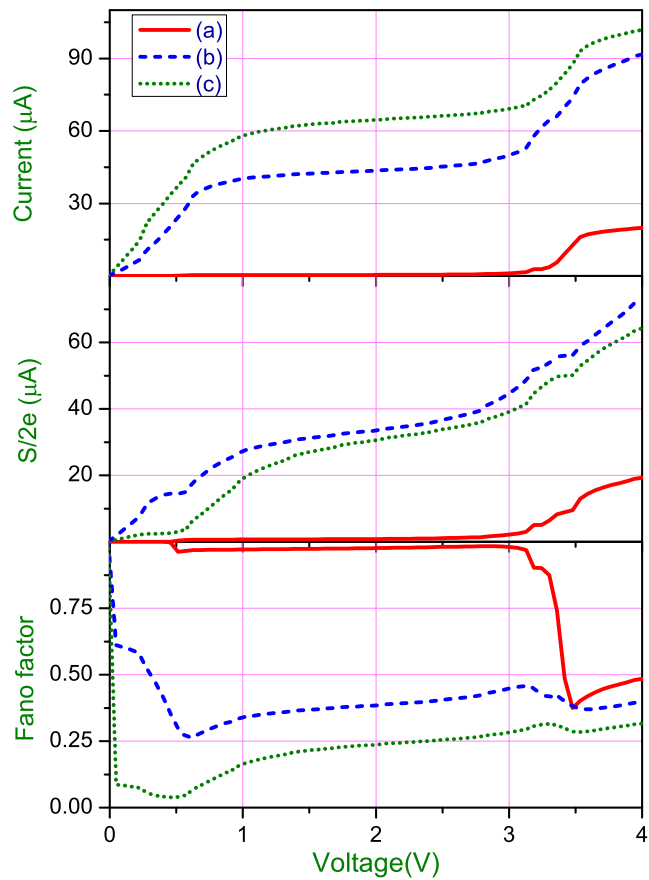}}
\caption{(color online.) Current(top panel), noise power (middle panel) and Fano factor(bottom panel) versus
applied bias voltage of the CNT/aGNR/CNT junction in the absence of magnetic field for different $a, b$ and $c$ cases.}
\label{p3}
\end{figure}

In Fig. \ref{p2}, we illustrate the logarithmic scale of transmission function versus energy of
the CNT/aGNR/CNT junction in the absence of magnetic field for $a, b$ and $c$ cases (see Fig.\ref{p1}).
As shown in Fig.\ref{p2}, for an electron with energy $E$, comes from the left electrode, the probability of
transmission function reaches its saturated value (resonance peaks) for the particular energy values. These resonance
peaks are related to the eigenenergies of the individual aGNR. When the electrons travel from left CNT electrode to the right one through the aGNR,
the electron waves propagating along the two branches of aGNR may suffer a relative phase shift between
themselves. Consequently, there might be constructive or destructive interference due to the superposition
of the electronic wave functions along the various pathways. Therefore, the transmission probability will change.
We observe some anti-resonant states appear in the transmission probability in the $a$ and $b$ cases.
In case $a$ there are two types of QI effects known as Fano-resonance (asymmetric line shape) about $1.6<E<1.7$ and anti-resonance (symmetric line shape) about $E_F$, while in case $b$ the Fano-resonance only exists. Critical difference between these QI effects is that a Fano-resonance does require a localized state at the position of the asymmetric peak, however an anti-resonance does not require a localized state at the position of a negative peak.
To obtain a deeper insight into the electron transport through the CNT/aGNR/CNT junction, we have plotted the
current, noise power and Fano factor as a function of applied voltage for the three $a, b$ and $c$ cases in the absence of magnetic field in Fig. \ref{p3}.
An applied voltage shifts the chemical potentials of two CNT electrodes relative to each other by $eV$, with $e$ the electronic charge.
When the aGNR energy level is positioned within such bias window, current will flow. For the same voltage $V$ the current
amplitude of the three $a, b$ and $c$ cases exhibit different response to the applied bias voltage $V$.
This is due to the QI effects of the electron waves traversing through the different branches of the aGNR.
It is observed that the $I-V$ curves shows staircase-like structure which indicates that a new channel is opened.
 With the increase of the bias voltage $V$, the electrochemical potentials on the
electrodes are shifted gradually, and finally cross one of the quantized energy levels of the aGNR.
Therefore, a current channel is opened up and provides a jump in the $I-V$ characteristic.
Apart from different current amplitude that three $a, b$ and $c$ cases show, they also show different threshold voltage
$V_t$ in which the current appears. Indeed, two $b$ and $c$ cases have shown
almost same threshold voltage $V_t$ which is much more smaller than case $a$.
\begin{figure}
\centerline{\includegraphics[width=0.55\columnwidth]{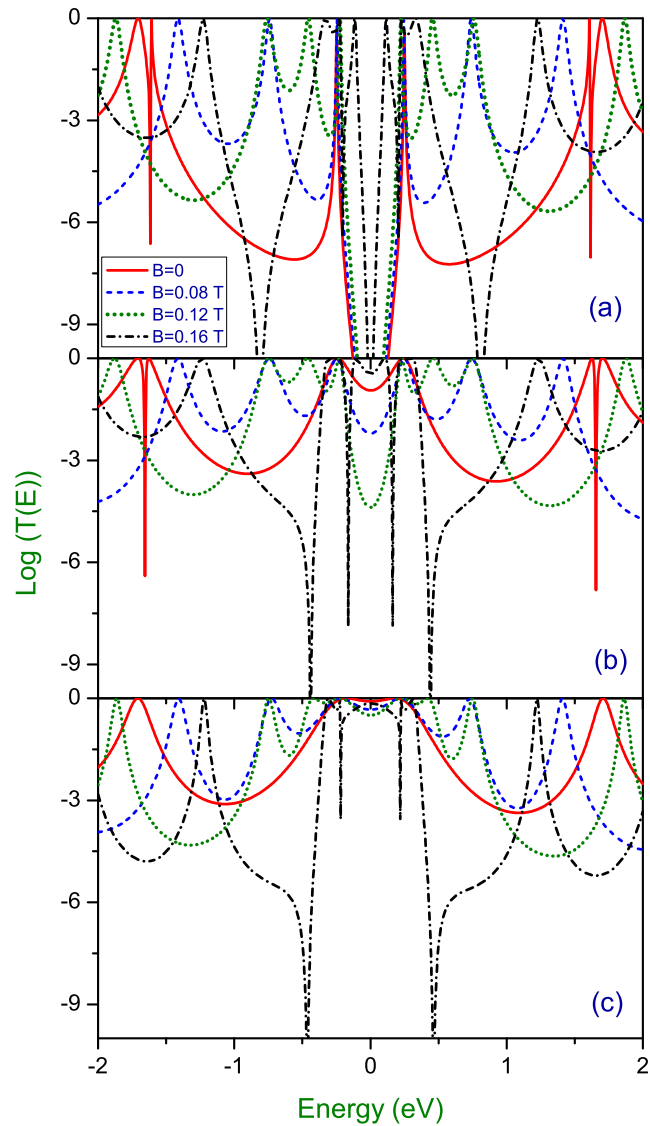}}
\caption{(color online.) The logarithmic scale of transmission probability as a function of the injecting electron
energy $E$ of the CNT/aGNR/CNT junction for different $a, b$ and $c$ cases for different magnetic
fields strength $B=0, 0.08, 0.12$ and $0.16 T$. }
\label{p4}
\end{figure}

\begin{figure}
\centerline{\includegraphics[width=0.5\columnwidth]{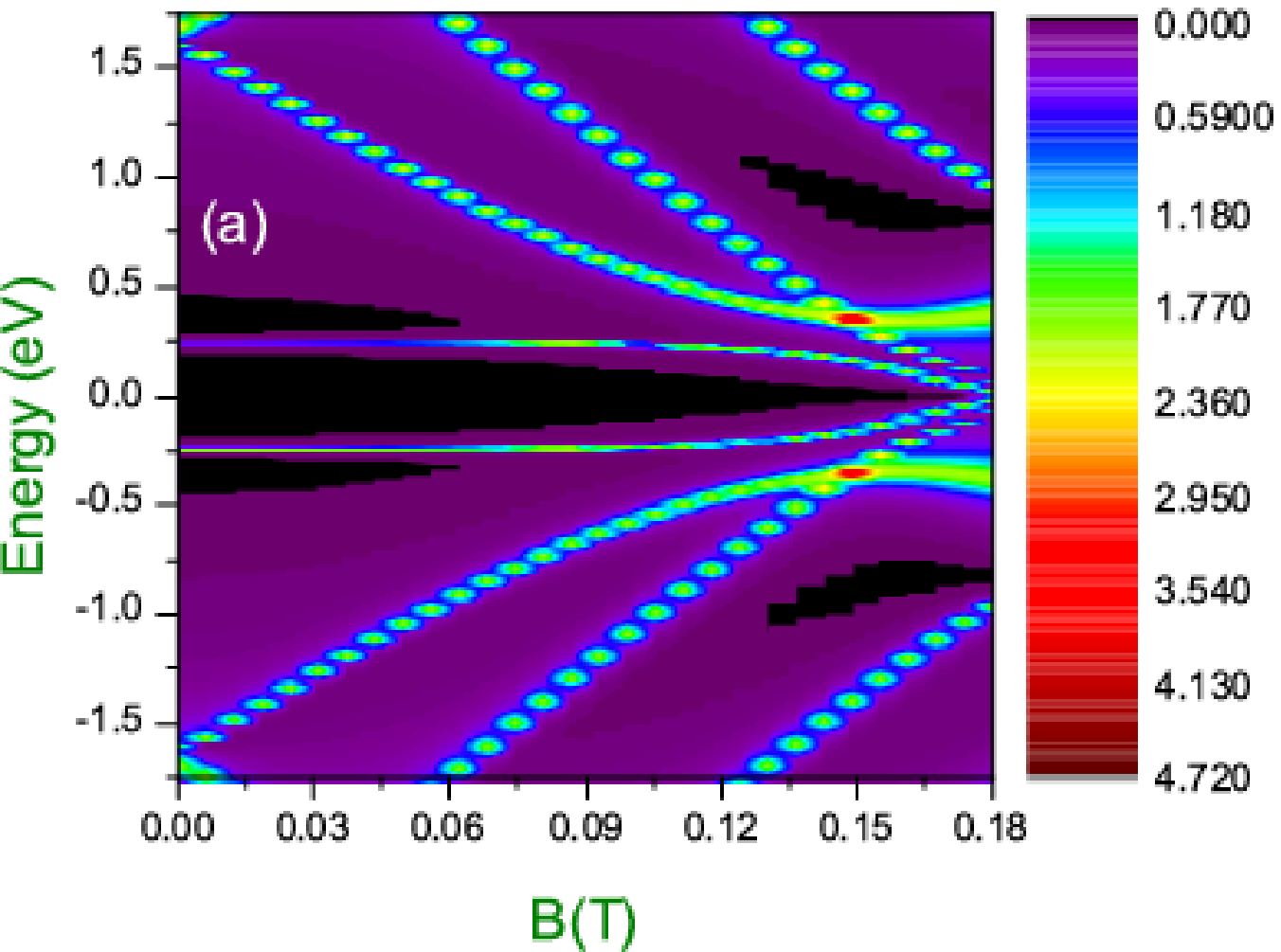}}
\centerline{\includegraphics[width=0.5\columnwidth]{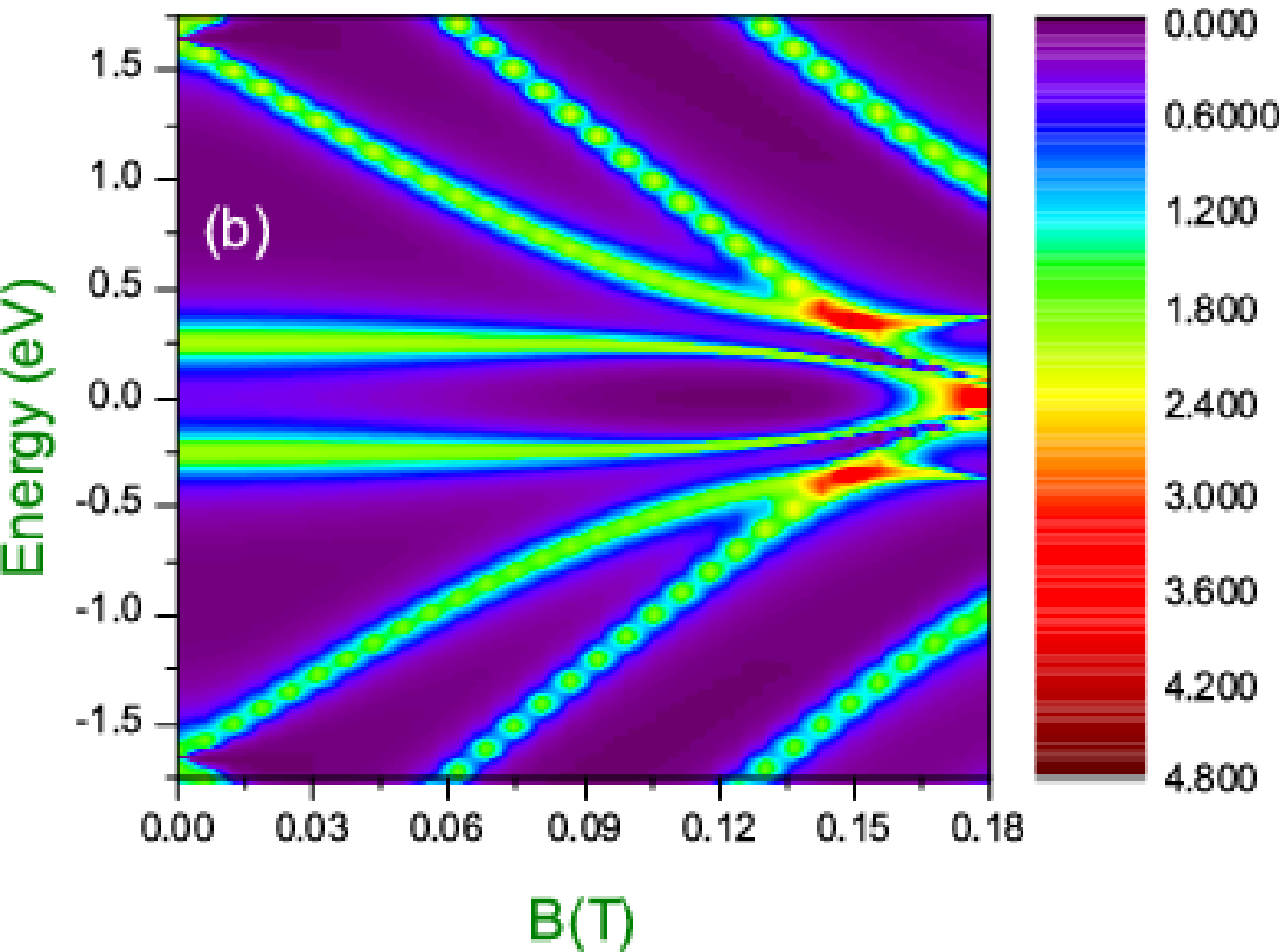}}
\centerline{\includegraphics[width=0.5\columnwidth]{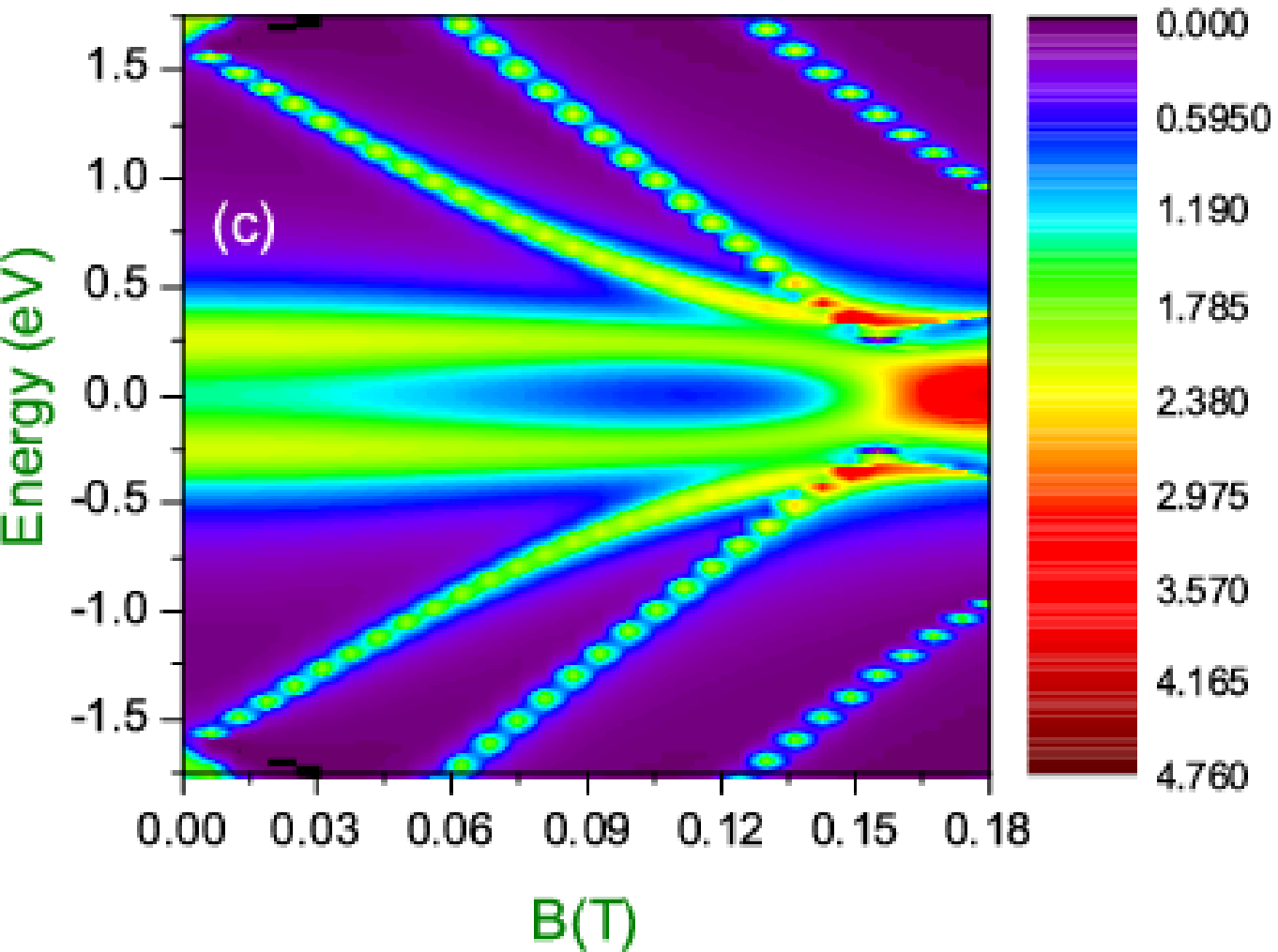}}
\caption{(color online.) The spectrum of local density of states of CNT/aGNR contact position as a function of the injecting electron
energy $E$ and magnetic field $B$ for different $a, b$ and $c$ cases.}
\label{p5}
\end{figure}
We have also depicted the noise power of current fluctuations, middle panel of Fig. \ref{p3}, for the three $a, b$ and $c$ cases as a function
of the bias voltage. It is apparent that the noise power and current have analogous step characteristics
with increasing bias-voltage. The noise power in the case $a$ has least magnitude in compare with the two $b$ and $c$ cases.
The one point which  should also be addressed is that though the general behaviour of the noise power and current are the same, but cases $b$ and
$c$ changes their relative trend in response to the bias-voltage. The bottom panel of Fig. \ref{p3} shows the bias-voltage dependent Fano factor for the CNT/aGNR/CNT system. It should be noticed that we neglect all the interactions between current carriers and electron correlations are associated only with the Pauli principle. After the first step in $I-V$ curve, the shot noise goes from the Poisson limit $(F=1)$ to Sub-Poisson
region $(F<1)$ for the case $a$. This demonstrates that the electrons are correlated after the tunneling process. This behavior does not exist for
two $b$ and $c$ cases since the shot noise already attains the Sub-Poisson region.

\par
\begin{figure}
\centerline{\includegraphics[width=0.6\columnwidth]{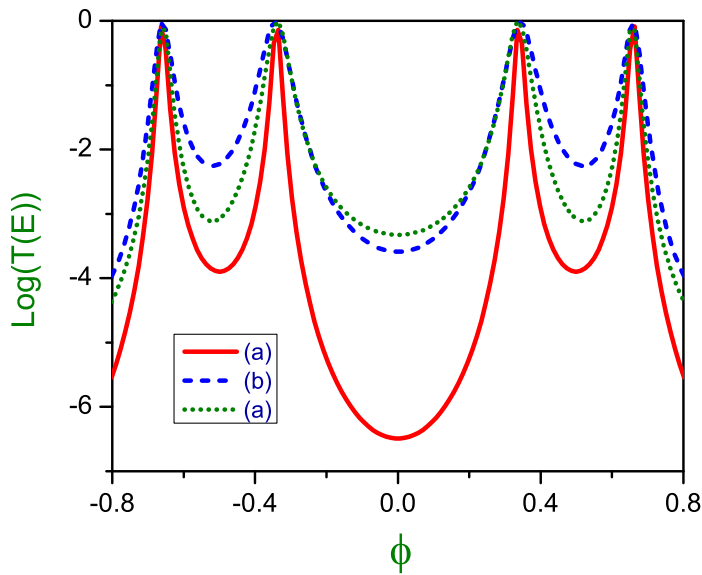}}
\caption{(color online.) The logarithmic scale of transmission probability as a function of magnetic flux for injection energy
of the tunneling electron $E=1 eV$ for different $a, b$ and $c$ cases. }
\label{p7}
\end{figure}

In what follows, we will apply an external magnetic field perpendicular to the aGNR and check its effect
on the transport properties of the three $a, b$ and $c$ cases. In Fig. \ref{p4}, we show logarithmic scale
of transmission function versus energy of the CNT/aGNR/CNT junction in the presence of magnetic field $B$
for the different $a, b$ and $c$ cases. A remarkable feature which happens after applying magnetic field
is that band gap starts to change by increasing magnetic field. This is due to variation of the hopping
integral $\tilde{t}$ in the presence of magnetic field. In all the three $a, b$ and $c$ cases, tuning magnetic
field strength brings different trend in the transmission function. For example in case $a$, by increasing
magnetic field strength anti-resonance at Fermi energy shows very sharp fall due to decreasing band gap.
It can be exploit to have high thermopower $S=-\frac{\pi^2K^2_BT}{3e}\frac{T'(E_F)}{T(E_F)}$ (measures the
logarithmic first derivative of the transmission function at Fermi energy). Apart from the anti-resonance
at Fermi energy, one can also see two other anti-resonances about energy $E\simeq\pm1eV$  for the presented
window of energy. As already mentioned above, this type of QI effects does not need local density of state.
In cases $b$ and $c$, we haven't seen any anti-resonance type of QI effects, while by increasing magnetic
field strength some Fano-resonance appears. These QI effects take place for magnetic field bigger than
$B>_\sim 0.13T$. In Fig.\ref{p5}, we present local density of state(LDOS) of CNT/aGNR contact position as function
of the injecting electron energy $E$ and magnetic field $B$ for different $a, b$ and $c$ cases. Fig. \ref{p5}(a) clearly
shows some off state region (black islands) which give rise to anti-resonance QI effects. As the magnetic field grows, off state
region around Fermi energy gradually starts to narrow. This behaviour reveals the transmissions trend around
the Fermi energy for case $a$. For higher magnetic field the off state region at Fermi energy shows robust
trend and exists for a whole cycle of magnetic flux (see Fig. \ref{p7}(. Moreover, for higher magnetic field
one can also see two off state region symmetric respect to the injected energy which can give rise to some
new anti-resonances. One more thing should be addressed here is that about $B\simeq0.14T$ and $E\simeq\pm0.4eV$
there are two regions with high LDOS(hot-red) which can introduce Fano-resonance in case $a$. In cases $b$
and $c$, LDOS spectrum dose not show any off state region and it is in well agreement with transmission
function which dose not  show any anti-resonances. In addition, the signature of Fano-resonances are couple
with high LDOS (hot-red) regions that clearly appear for $B>_\sim 0.13T$.

\begin{figure*}
\centerline{\includegraphics[width=1.\columnwidth]{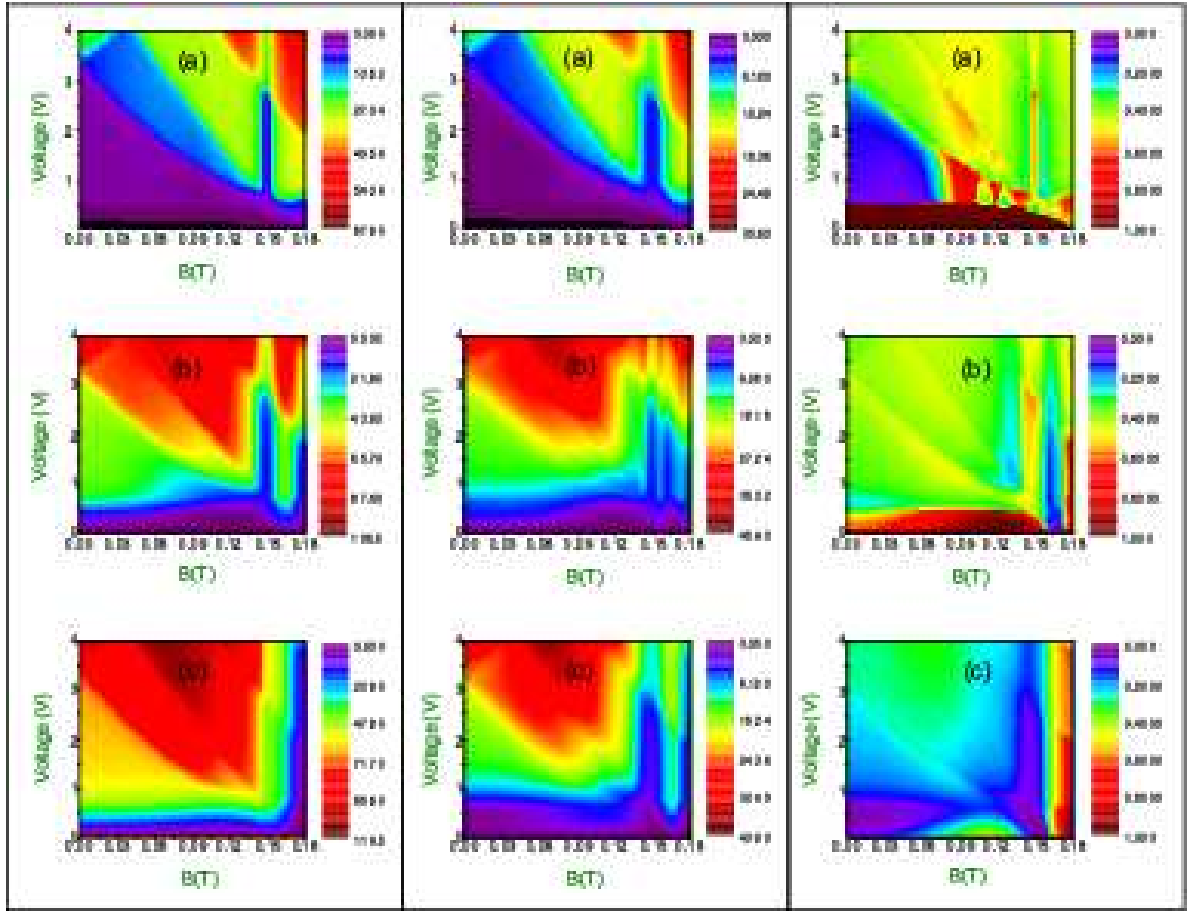}}
\caption{(color online.) Current(left panel), noise power (middle panel) and Fano factor(right panel) versus
applied bias voltage and magnetic field $B$ of the CNT/aGNR/CNT junction for different $a, b$ and $c$ cases.}
\label{p6}
\end{figure*}

Now for further clarification in Fig. \ref{p6}, we have illustrated current, noise power and Fano factor as function of
applied bias voltage and magnetic field $B$  for different $a, b$ and $c$ cases. Both current and noise power show almost
same trend in respect to the bias voltage and magnetic field. In case $a$,  presence of a threshold voltage $V_t$ (dark-black region )is clear
and it goes to disappear by growing magnetic field strength which can be referred as the band gap narrowing.
 This threshold voltage dose not exist in two $b$ and $c$ cases in which current
follows immediate after turning bias voltage and finds it saturation value for higher voltage. Comparing current density plots of three $a, b$
and $c$ cases also shows that for presented window of bias voltage and magnetic field, case $c$ finds its saturation value for broader region.

As it can be seen from the noise power for case $a$ in the presence of applied magnetic field, the noise power
goes from the Poisson limit $(F=1)$ to Sub-Poisson region $(F<1)$. This transition takes place in different bias voltages
 due to the decreasing threshold voltage $V_t$. Case $b$ shows Poisson limit $(F=1)$ for very tiny region about $B\simeq0.1 T$ and
  low bias, elsewhere shows the Sub-Poisson limit. In case $c$, the noise power for different magnetic field
 strengths lies in the Sub-Poisson limit.

\par
In Fig. \ref{p7}, we have shown the AB oscillations in the logarithmic scale of transmission probability as a
 function of magnetic flux ($\phi$) for injection energy of the tunneling electron $E=1 eV$ for the different $a, b$ and $c$ cases.
 Transmission probability shows periodic pattern with a period of $\phi_0$. This periodicity is
 seen for all energies of tunneling electron which is not presented here.
\section{Summary and conclusion}
In summary, we have numerically investigated the coherent electron transport through armchair graphene nanoribbon (aGNR)
sandwiched between carbon nanotube (CNT) electrodes as CNT/aGNR/CNT system. Using the generalized Green's function technique
based on tight-binding model and the Landauer-B\"uttiker formalism, we have calculated
the transmission probability, current, noise power and Fano factor in the \textbf{\emph{presence}} of magnetic field. We have also considered
three different contact positions cases(as labeled $a, b$ and $c$) in order to check the existence of quantum interference(QI) (see Fig. \ref{p1}).
Our calculations show the presences of two type of QI effects which can lead to appear anti-resonance and Fano-resonance \textbf{\emph{peaks}} in the transmission probability which can affect on the transport properties of the CNT/aGNR/CNT system. It is found that all the three $a, b$ and $c$ cases display different current and noise power amplitudes. Moreover, they show different threshold voltages $V_t$ in which the current and noise power occurs. Indeed, two $b$ and $c$ cases show almost same threshold voltages $V_t$ which are much more smaller than case $a$.
 
We have also considered two QI sources, contact positions and magnetic field, to check their mutual effects on the coherent transport characteristics of CNT/aGNR/CNT system. Our calculations unveil that in all the three $a, b$ and $c$ cases by changing magnetic field, the current amplitude and threshold voltage $V_t$ starts to change. Our results also show that the shot noise characteristic, in the Poisson limit ($F=1$) or sub-Poisson limit ($F<1$), and the maximum value of the Fano factor \textbf{\emph{strongly}} depend on contact positions and the magnetic field strength. Thus we can emphasize that the transport properties of the CNT/aGNR/CNT system such as noise power and Fano factor can be controlled very significantly by tuning the QI effects.
These phenomena can be utilized in designing the future nano-electronic devices.


\end{document}